\documentclass[twocolumn,showpacs,preprintnumbers,amsmath,amssymb,floatfix]{revtex4}
\usepackage{graphicx}
\usepackage{color}

\begin{document}

\title{The origin of hysteresis in resistive switching in magnetite is Joule heating.}

\author{A.~A.~Fursina$^{1}$, R.~G.~S.~Sofin$^{2}$, I.~V.~Shvets$^{2}$, D.~Natelson$^{3, 4}$}

\affiliation{$^{1}$ Department of Chemistry, Rice University, 6100 Main St., Houston, TX 77005}
\affiliation{$^{2}$ CRANN, School of Physics, Trinity College, Dublin 2, Ireland}
\affiliation{$^{3}$ Department of Physics and Astronomy, Rice University, 6100 Main St., Houston, TX 77005}
\affiliation{$^{4}$ Department of Electrical and Computer Engineering, Rice University, 6100 Main St,.Houston, TX 77005}

\date{\today}


\begin{abstract} 
In many transition metal oxides the electrical resistance is observed
to undergo dramatic changes induced by large biases.  In magnetite,
Fe$_3$O$_4$, below the Verwey temperature, an electric field driven
transition to a state of lower resistance was recently found, with
hysteretic current-voltage response.  We report the results of pulsed
electrical conduction measurements in epitaxial magnetite thin films.
We show that while the high- to low-resistance transition is driven by
electric field, the hysteresis observed in $I-V$ curves results from
Joule heating in the low resistance state.  The shape of the
hysteresis loop depends on pulse parameters, and reduces to a
hysteresis-free ``jump'' of the current provided thermal relaxation is
rapid compared to the time between voltage pulses.  A simple
relaxation time thermal model is proposed that captures the essentials
of the hysteresis mechanism.
\end{abstract}

\maketitle


Dramatic changes in resistance induced by electric fields, so called resistive switching (RS), have recently attracted much attention due to this phenomenon's potential application in memory devices (resistive random access memory, ReRAM) \cite{2008_Sawa_review, 2007_Aono_review}. RS from high- to low-resistance states is driven by application of high voltage, and corresponding up-and-down sweeps of current-voltage ($I$-$V$) characteristics often show hysteresis, {\it i.e.} in sweeps up and down in bias voltage, the current does not retrace itself. Systems exhibiting hysteretic RS include organic compounds \cite{2007_APL_organic_RS} and transition-metal oxides such as widely-studied colossal resistance manganites \cite{1997_Tokura_first}, perovskites ({\it e.g.} SrTiO$_3$ \cite{2006_Nature_SrTiO3_Waser}), 1D cuprates Sr$_2$CuO$_3$ \cite{2000_PRB_Tokura_cuprate}, NiO \cite{2006_APL_NiO_RS}, TiO$_2$ \cite{2005_JAP_TiO2_Waser} {\it etc}.

For some RS systems, while sweeping out a hysteresis loop in $I$-$V$ with a switch to a low resistance state at high bias, the low resistance state persists down to zero current as voltage approaches zero. This behavior is often the case for RS systems where the switching is based on metallic filament formation at a transition point \cite{2006_Nature_SrTiO3_Waser}. However, for some RS systems the low-resistance state persists only in some voltage interval, and the system returns to the high-resistance state before voltage returns to zero.  This is the case for some complex oxides\cite{1997_Tokura_first,2008_Sun_APL,2008_Li_EPL} as well as for magnetite nanostructures, which were recently shown to exhibit RS at low temperatures \cite{Our_magnetite_2007,2008_APL_HAR_Cr}.

Magnetite, Fe$_3$O$_4$, is an example of strongly correlated material. In equilibrium, bulk magnetite undergoes a structural transition at the Verwey temperature, $T_{V}\sim$120~K, accompanied by three-order-of-magnitude change in electrical conductivity, {\it i.e.} a metal-insulator transition (MIT) \cite{1939_Verwey_first_Nature}.  Recently  we demonstrated that magnetite nanoparticles and thin films, once in the insulating state below $T_{V}$, exhibit RS under a sufficiently large voltage bias \cite{ Our_magnetite_2007}.  By examining RS systematically in different device geometries, the switching was demonstrated to be driven by the applied in-plane electric field.  This is in contrast to previously observed transitions in magnetite driven by Joule heating of the samples above $T_{V}$ under bias \cite{1969_Tinduced_MIT_Fe3O4_1, 1969_Tinduced_MIT_Fe3O4_2}.  When the voltage is swept continuously, the electric field-driven switching takes place at either polarity of voltage with well-pronounced hysteresis.

In this paper we determine the origin of hysteresis in magnetite structures based on epitaxial thin films.  Through extensive voltage pulse measurements with controlled pulse parameters we unambiguously show that, while the high resistance to low resistance switching is driven by electric field, the hysteretic behavior originates from local Joule heating of the channel once the system is switched to the low-resistance state.  A very simple thermal model agrees with the data with appropriately chosen parameters.  The parameter values required to achieve quantitative consistency with the data demonstrate that one must consider heating and thermal transport beyond just the magnetite film itself.

\begin{figure}[t]
\begin{center}
\includegraphics[clip,width=8cm]{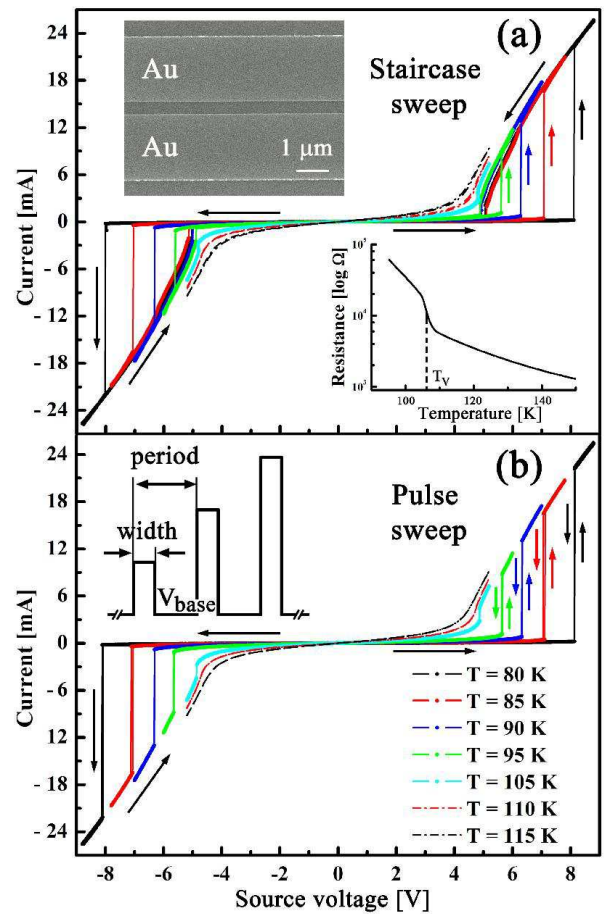}
\end{center}
\vspace{-3mm}
\caption{\small Typical $I$-$V$ curves at different temperatures in continuous staircase (a) and pulse (b) sweep modes. Arrows indicate the direction of voltage sweeps. Above a certain temperature (in this case $\sim$ 105~K) sharp jumps in current are not observable. Top (a) inset shows a scanning electron micrograph of a two-terminal device on the surface of a magnetite film; bottom (a) inset is a temperature dependence of resistance demonstrating Verwey transition at $T_{V}\sim$ 108~K.  The (b) inset schematically represents the pulse parameters for the pulsed sweeps.  Successive applied voltages differ by 1-15~mV depending on the total sweep range.}
\label{fig1}
\vspace{-3mm}
\end{figure}

Epitaxial magnetite thin films with a 50~nm thickness  were grown on $\langle$100$\rangle$ MgO single crystal substrates by oxygen-plasma-assisted molecular beam epitaxy.  Details of the growth process have been reported elsewhere \cite{2008_Shvets_growth, 2004_JAP_Schvets}. The films were characterized by reflection high-energy electron diffraction ({\it in situ}, during growing), high resolution X-ray diffraction measurements, Raman spectroscopy and resistance measurement to prove crystalline quality and stoichiometry of the samples \cite{Shvets_backscat,Shvets_high_res_Xray}.  The films show a jump in the temperature dependence of resistance at $T \sim$ 108~K (Fig.~\ref{fig1}a inset), characteristic of the Verwey transition in magnetite thin films.

Devices for two-terminal measurements were prepared by electron beam lithography.  A channel length 190 - 900~nm is defined by two 5 - 10~$\mu$m wide current leads (Fig.~\ref{fig1}a inset) which are connected to micrometer-size pads (300 $\times$ 300 $\mu$m).  Upon testing several different contact metals (Au, Pt, Cu, Fe and Al), copper showed the lowest contact resistance with the magnetite film. Thus, the electrodes were made of 6~nm Cu / 15~nm Au thin layers deposited by electron beam evaporation.


Electrical characterization of the samples was performed by a standard two-terminal method using a semiconductor parameter analyzer (Hewlett Packard 4155A).  The voltage was applied to the source lead with the drain grounded, and current flowing through the channel was monitored.  Two different measurement methods were used:  a continuous staircase sweep of the source voltage up and down; and a pulsed sweep implemented with controlled pulse parameters.  Each voltage sweep consists of 1000 points equally spaced in voltage.  Measurements were performed in the temperature range 80-300~K with 5~K steps.  

At high temperatures $I$-$V$ curves have a slightly non-linear shape (Fig.~\ref{fig1}a,~b) typical for many transition metal oxides.  Below a certain temperature, coincident with the Verwey temperature inferred from measurements of the zero-bias resistance, a sharp jump in current was observed as the source voltage reached a critical value, as described previously \cite{Our_magnetite_2007}. 
At 80~K, for example, the current changes abruptly by a factor of 100 at a particular switching voltage.  The state after transition (``On'' state) shows an approximately linear $I$-$V$ dependence with a much smaller differential resistance than that of high-resistance (``Off'') state prior to the transition.  At a given temperature and device geometry, the switching happens at a critical voltage, $V^{On}_{sw}$, and as the voltage is swept back the system remains in the On state until it switches back to the Off state at a switch-off voltage, $V^{Off}_{sw}$.   For a continuous staircase sweep $V^{Off}_{sw}$ is always lower than $V^{On}_{sw}$, resulting in well defined hysteresis (Fig.~\ref{fig1}a). The transition is symmetrical along $V$-axis, with identical transitions occurring at positive and negative voltage sweeps.

Switching phenomena are observed at temperatures right below Verwey temperature ($\sim$108~K) and, thus, magnetite is comparatively insulating in the undisturbed state (see zero-bias $R$ vs $T$ dependence in Fig.~\ref{fig1}a inset).  Previously we demonstrated that the observed transition is driven by electric field, not by thermal heating of the magnetite sample above Verwey temperature \cite{Our_magnetite_2007}.  In the high resistance state ($|V| < V^{On}_{sw}$) heating of the sample is comparatively negligible, whereas in the On state dissipated power significantly increases, making Joule heating of the sample much more likely.  Below we use pulsed measurements to demonstrate that this On state heating is the origin of the apparent hysteretic behavior in our system. 

We performed pulsed sweep experiments, with controlled duration of the voltage pulse (pulse width), the time between consecutive pulses (pulse period) and the voltage value at rest between the pulses (base voltage) (Fig.~\ref{fig1}b inset). In these pulsed measurements the base voltage may be kept $V_{base}=0$~V, eliminating Joule heating and allowing cooling of the channel between pulses.  The dependence of system response on pulse width and pulse period can indicate heating dynamics of the sample (especially in the On state).

\begin{figure}[t]
\begin{center}
\includegraphics[clip,width=8cm]{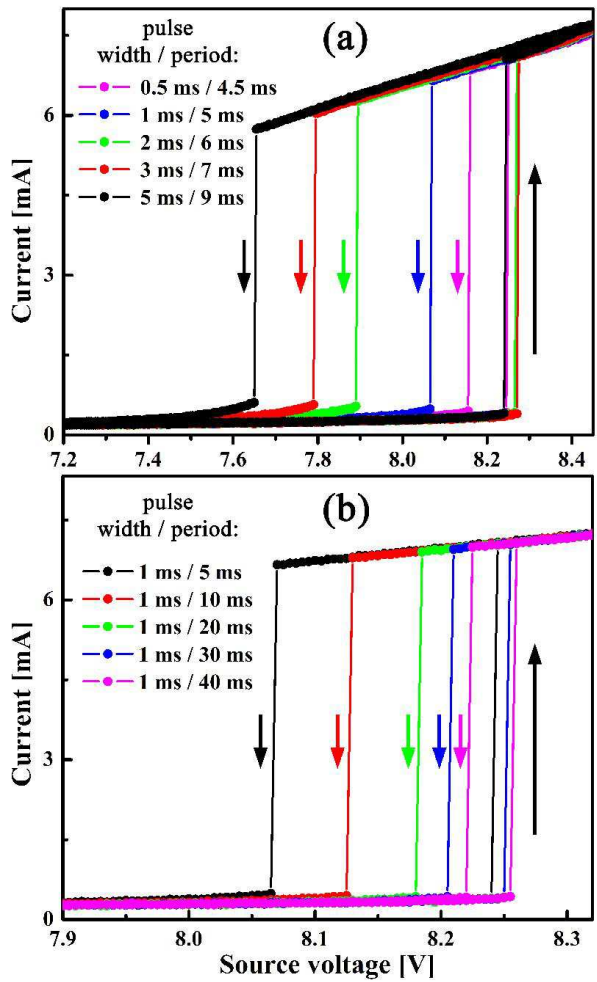}
\end{center}
\vspace{-3mm}
\caption{\small Dependence of hysteresis ($I$-$V$ curve) shape at 80K in pulse sweep mode on (a) pulse width while the time sample rests at $V_{base}$ is kept constant (4~ms) and (b) pulse period, while pulse width is kept the same (1~ms). Base voltage, $V_{base}$, is always set to 0 V.}
\label{fig2}
\vspace{-3mm}
\end{figure}

First, we investigated the system response to pulse sweeps with different {\it widths}: we varied the pulse duration while keeping fixed the time system stays at base voltage (pulse period - pulse width = const, ($V_{base} = 0$~V), see fig.~\ref{fig1}b inset). Fig.~\ref{fig2}a shows resultant $I$-$V$ curves (only positive voltages for better visualization) at 80~K. The shape of the hysteresis clearly changes as pulse width increases:  $V^{Off}_{sw}$ moves farther away from $V^{On}_{sw}$, which remains independent of the pulse width (Fig.~\ref{fig2}a). Note that the current is consistently higher for longer pulses.

In a second set of pulse experiments we made voltage sweeps with various pulse {\it periods} and constant pulse width. The shortest possible pulse width (0.5~ms in our setup) was used to minimize heating of the sample. The increase of pulse period has an opposite effect on $V^{Off}_{sw}$ position than the increase of pulse width: $V^{Off}_{sw}$ moves closer to $V^{On}_{sw}$ as the pulse period increases (Fig.~\ref{fig2}b), {\it i.e.} as the system rests at zero bias voltage between pulses for longer times. Note, again, that $V^{On}_{sw}$ position remains independent of pulse period. At long enough period times ($>$ 100 ms) $V^{On}_{sw}$ and $V^{Off}_{sw}$ are the same value.  Fig.~\ref{fig1}b shows $I$-$V$ curves at different temperatures obtained by pulse sweeps with a 100~ms pulse period.  Instead of hystereses there are sharp jumps in On state and back to Off state at almost the same voltage: $V^{Off}_{sw} \approx V^{On}_{sw}$.  

\begin{figure}[t]
\begin{center}
\includegraphics[clip,width=8cm]{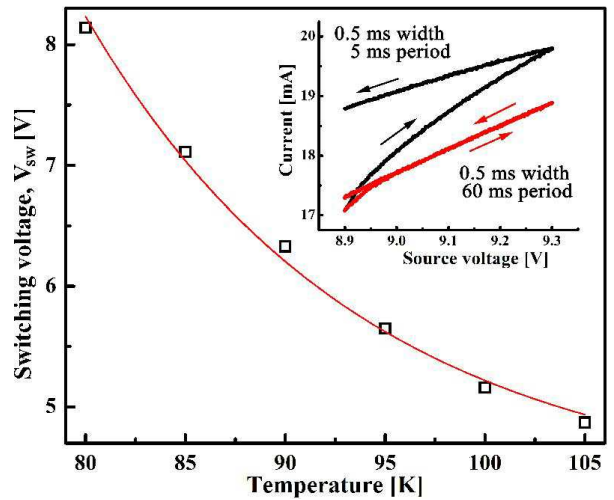}
\end{center}
\vspace{-3mm}
\caption{\small Temperature dependence of switching voltage (open squares) and its exponential fit (solid line). The inset shows $I$-$V$ curves in the On state in pulse sweep mode with the same pulse width (0.5 ms) but different pulse periods, showing that rapid pulse repitition leads to more severe heating, as expected.}
\label{fig3}
\vspace{-3mm}
\end{figure}

While hysteretic RS (with differentiated $V^{On}_{sw}$ and $V^{Off}_{sw}$ values) was previously observed in many systems, here we demonstrate that in magnetite there is a single \textit{intrinsic} switching value, $V_{sw}=V^{On}_{sw}=V^{Off}_{sw}$, and  no intrinsic hysteresis.  Thus, the switching voltage is a characteristic parameter of the system: at every temperature and device geometry, there is a certain switching voltage required to drive the transition.  This $V_{sw}$ decreases as temperature approaches $T_{V}$, following a nearly exponential dependence (Fig.~\ref{fig3}).  The separation of $V^{On}_{sw}$ and $V^{Off}_{sw}$ values, shifting of $V^{Off}_{sw}$ position in various pulse period and width sweeps, and the observation of hysteresis are easily explained by heating of the sample in the On state.  

We first give a qualitative picture of how On state heating leads to hysteresis, and then provide a quantitative description with a simple model of channel temperature during a pulse sweep.  The sample begins at an initial temperature $T_{set}$; for this temperature and a given channel length, there is a corresponding switching voltage, $V_{sw}(T_{set})$ (Fig.~\ref{fig3}).  Consider performing a sweep with pulse parameters selected to give hysteresis.  First, as we start sweeping voltage from 0~V up, the system remains approximately at $T_{set}$ prior to the transition point due to the comparatively small amount of Joule heating.  After application of a pulse with $V > V_{sw}(T_{set})$, the system is driven into the On state.  With the resulting increased conductance (and, thus, current) the Joule heating of the channel is much larger, elevating the local effective temperature of the channel to $T_{cur} > T_{set}$.  Non-steady-state warming of the sample continues further in the On state.  This is clear from the non-linear increase of the current and the fact that the $I$-$V$ trace in the On state does not retrace itself (Fig.~\ref{fig3} inset).  Note that sufficiently increasing the pulse period does allow $I$-$V$ curves to retrace themselves on up and down sweeps (Fig.~\ref{fig3} inset).  As the pulse voltage is swept back below $V_{sw}(T_{set})$, the channel remains in the On state because the channel is at the elevated effective temperature, $T_{cur}$,  and the switching voltage is therefore lower: $V_{sw}(T_{cur}) < V_{sw}(T_{set})$.   


If the channel cools down to $T_{set}$ between two consecutive pulses, the system remain in the Off state for any voltage applied that is lower than $V_{sw}(T_{set})$ and jumps are observed instead of hysteresis (Fig.~\ref{fig1}b).  If the pulse period is short compared to the thermal relaxation time, then the channel remains at $T_{cur} > T_{set}$, and we observe the separation of $V^{On}_{sw} = V_{sw}(T_{set})$ and $V^{Off}_{sw}=V_{sw}(T_{cur})$, and resulting hysteresis.  Upon increasing the pulse period (system stays longer at $V_{base} = 0$~V), the channel has more time to cool after application of previous pulses, $T_{cur}$ is closer to $T_{set}$ and $V^{Off}_{sw}$ moves closer to $V^{On}_{sw}$ in excellent agreement with the experimental data (Fig.~\ref{fig2}b).  
In experiments on increasing pulse widths, it is clear that longer pulses heat the channel more than shorter pulses, thus $T_{cur}$ is higher in the former case, corresponding to lower $V_{sw}(T_{cur})$. Thus, experimentally observed moving $V^{Off}_{sw}$ away from $V^{On}_{sw}$ as the pulse width increases (Fig.~\ref{fig2}a) is completely consistent with this heating picture.


As qualitatively described above, $V^{Off}_{sw}$ position is related to the current local temperature of the channel.  To quantitatively describe the movement of $V^{Off}_{sw}$ position in sweeps with different pulse parameters, we should consider a model to estimate the effective channel temperature as a function of applied pulsed voltage (and time).  Detailed thermal modeling is very challenging because of the highly local character of the heating and the difficulty of capturing all the relevant heat transfer processes in these nanostructures.  Instead we consider a simple model where the return to equilibrium state (relaxation) is described by a relaxation time, $\tau$ \cite{heat_trasf_book}, and the temperature change is:
\begin{equation}
\frac{\partial T}{\partial t}(relaxation)=-\frac{T-T_{set}}{\tau}
\label{eq1}
\end{equation}
where $T$ is a current temperature of the channel.  The heat flux into the sample is obviously $dQ_{in}/dt=V \times I$.  We further assume the heat dissipation can be simply described as 
\begin{equation}
\frac{dQ_{out}}{dt}=-C_{v} \frac{T-T_{set}}{\tau},
\label{eq2}
\end{equation}
where $C_{v}$ is a heat capacity of the sample (in Joule/K).

Note that this model assumes $C_{v}$ and $\tau$ to be independent of temperature, which is clearly an idealization.  The temperature variation with respect to time, $\partial T/\partial t$, considering total heat flux through the system, {\it i.e.} $\frac{\partial}{\partial t} (Q_{in}+Q_{out})$, is given by:
\begin{equation}
C_{v} \frac{\partial T}{\partial t}=\frac{\partial}{\partial t} (Q_{in}+Q_{out})=V \times I -C_{v} \frac{T-T_{set}}{\tau}.
\label{eq3}
\end{equation}
By solving Eq.~(\ref{eq3})
we derive an expression for a temperature, $T(t+dt)$, given the current temperature, $T(t)$, in a moment of time, $dt$:

$$ T(t+dt)=T_{set}+\frac{V \times I \times \tau}{C_{v}} \times (1- \exp{(-\frac{dt}{\tau})})+$$

\begin{equation}
+(T(t)-T_{set}) \times \exp{(-\frac{dt}{\tau})}
\label{eq4}
\end{equation}

When $V \ne 0$, both heating and cooling (by relaxation to equilibrium $T_{set}$) processes occur described by second and third terms of the right side of Eq.~(\ref{eq4}), respectively.  Between pulses ($V_{base}$=0~V) only heat dissipation takes place and Eq.~(\ref{eq4}) reduces to the expected exponential decay of the temperature with the relaxation time, $\tau$, as a scaling factor:
\begin{equation}
T(t+dt)=T_{set}+(T(t)-T_{set}) \times \exp{(-\frac{dt}{\tau})}
\label{eq5}
\end{equation}
There are two fitting parameters in this model that govern the evolution of the current temperature: heat capacity, $C_{v}$, and relaxation time, $\tau$. 

\begin{figure}[t]
\begin{center}
\includegraphics[clip,width=8cm]{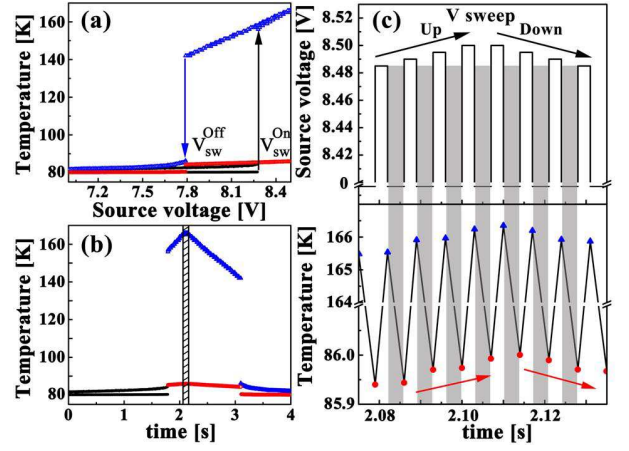}
\end{center}
\vspace{-3mm}
\caption{\small Calculated temperature in relaxation time model (see
  Eq.~(\ref{eq4})) with respect to source voltage (a) and
  corresponding time (b) for a 3~ms-wide / 7~ms-period voltage pulse
  sweep. Black points reflect $T$ values before the transition point,
  $V<V_{sw}$. After the transition point, blue triangles show $T$ immediately
  after pulse application and red circles show $T$ after relaxation
  between two sequential pulses.  The hatched area in (b) indicates the time
  interval depicted in detail in (c):  Applied pulsed voltage as a
  function of time (top) with corresponding calculated temperature of
  the channel (bottom).  Shaded areas indicate time intervals when system is
  is not under bias ({\it i.e.}  source voltage is zero) and only
  relaxation to $T_{set}$  is taking place (see Eq.~(\ref{eq5})).}
\label{fig4}
\vspace{-3mm}
\end{figure}

An example of $T$ variation calculated by Eq.~(\ref{eq4}) is shown in
Fig.~\ref{fig4} with real experimental data for sourced voltage, $V$,
and measured current, $I$, (3~ms-width, 7~ms-period pulse sweep) and
$C_{v}$ and $\tau$ values as will be discussed below.
Figs.~\ref{fig4}a, b show the calculated temperature of the channel as a
function of source voltage and corresponding time, respectively, over a large voltage
range around transition point.  Black points represent
$T$ values before the transition to the On state and they do not deviate
much from $T_{set}$ (80 K in this case). As the system is switched to the
ON state the temperature profile changes drastically: As pulse voltage
is sourced, $T$ increases and blue triangles show the calculated $T$
values right after pulse application (``high-$T$'' state).  After
relaxation for a time between pulses, the channel cools down to
temperatures represented by red circles (``low-$T$'' state) which are
only slightly higher than original $T_{set}$.  For better visualization,
Fig.~\ref{fig4}c zooms into the small region indicated by the hatched area
in Fig.~\ref{fig4}b at a turning point of a sweep (from sweep up to
sweep down in voltage).   It clearly demonstrates how the temperature first
rapidly rises as the pulse is applied and then decreases as the system
rests at $V_{base}=0 V$ (shaded regions), relaxing back toward $T_{set}$ =
80 K.  Both ``high-$T$'' and ``low-$T$'' temperatures follow the
applied voltage: $T$ increases as $V$ is swept up and starts
decreasing as sweep is reversed and $V$ decreases (follow the red
circles and blue rectangulars in Fig.~\ref{fig4}c bottom).

Each device has its defined single-valued $V_{sw}(T)$ function (an
example is shown in Fig.~\ref{fig3}) with $V_{sw}$ being a
characteristic parameter at a certain temperature.  During a voltage
sweep \textit{down}, the temperature at the moment the pulse is
applied decreases (follow the red circles in Fig.~\ref{fig4}c) while
$V_{sw}$ corresponding to this temperature increases.  Note that when
the pulse voltage is applied the relevant temperature is that of the
``low-$T$'' state (red circles in Fig.~\ref{fig4}c bottom), thermally
relaxed after application of the preceding pulse.  Switching from On
to Off states happens as soon as applied pulse $V$ value is lower than
$V_{sw}$ for the {\it current} temperature of the channel at the time
the pulse is applied.

\begin{figure}[t]
\begin{center}
\includegraphics[clip,width=8cm]{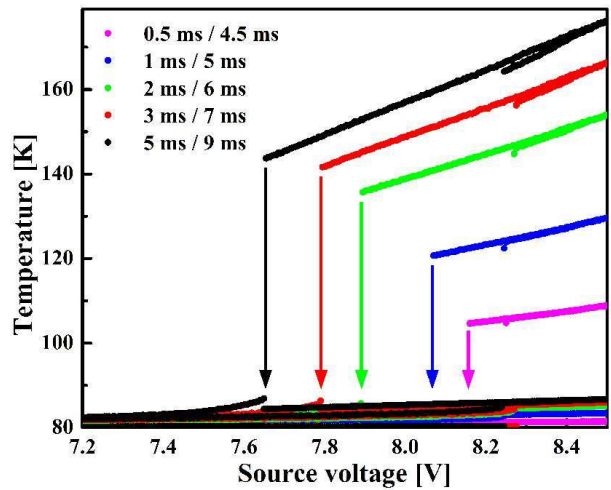}
\end{center}
\vspace{-3mm}
\caption{\small  The dependence of calculated temperature on source voltage in the relaxation time model for sweeps with different pulse widths and the same time between pulses (4 ms), $T_{set}$ = 80~K.  ``High-$T$'' (immediately after pulse application) and ``low-$T$'' (after relaxation between pulses) state temperatures are plotted in the same color, with the latter apparent at the bottom of the figure.  Arrows indicate switching back position for each sweep which happens when applied voltage becomes lower than $V_{sw}$ for a current temperature of the channel.}
\label{fig5}
\vspace{-3mm}
\end{figure}

We find that the proposed model can match experimental $V^{Off}_{sw}$ positions with appropriately chosen $C_{v}$ and $\tau$ values.  For each device at a given $T_{set}$ we can find a $C_{v}$ and $\tau$ pair which accurately describes the shifting of  $V^{Off}_{sw}$ position for sweeps with different pulse widths and the same time between pulses (pulse period - pulse width = const).  An example of temperature dependence on applied pulsed voltage is shown in Fig.~\ref{fig5} corresponding to experimental $I$-$V$ curves in Fig.~\ref{fig2}a.  The adjusted optimum values of $C_{v}$ and $\tau$ are 10$^{-6}$ Joule/K and 1.5$\times$10$^{-3}$ seconds, respectively.   

To get a sense of what these parameter values imply, we note that the estimated heat capacity value based on specific heat capacity and density of magnetite ($C_{v}\sim$ 40 Joule/(mol$\times$K) at 80~K \cite{1985_heat_capacit}, $\rho$ = 5.18 g/cm$^{3}$) and geometrical parameters of the channel is $\sim 10^{-13}$ Joule/K.  This is much lower than $C_{v}$ value required by the model to match the observed trends, strongly implying that heating is not confined only to the channel, but involves a much larger volume.  This also implies that the MgO substrate and Au electrodes are relevant to the thermal relaxation process. 

For sweeps with different pulse periods and the same pulse width, $V^{off}_{sw}$ positions can not be satisfactory modeled with a single $C_v$ and $\tau$ pair. For a fixed $C_v$, sweeps with longer pulse periods demand larger $\tau$ values to match the experimentally observed $V^{off}_{sw}$ positions. Longer period times imply  larger temperature variation over the time, relaxation process takes place (see Eq.~(\ref{eq5})).  This is almost certainly due to the failure of the idealized assumption of $C_{v}$ and $\tau$ being independent of temperature.  A much more sophisticated thermal model (incorporating the spatial temperature distribution and the temperature dependent thermal properties of the magnetite, the substrate, and the electrodes) is likely necessary for a complete quantitative picture.

\begin{figure}[t]
\begin{center}
\includegraphics[clip,width=8cm]{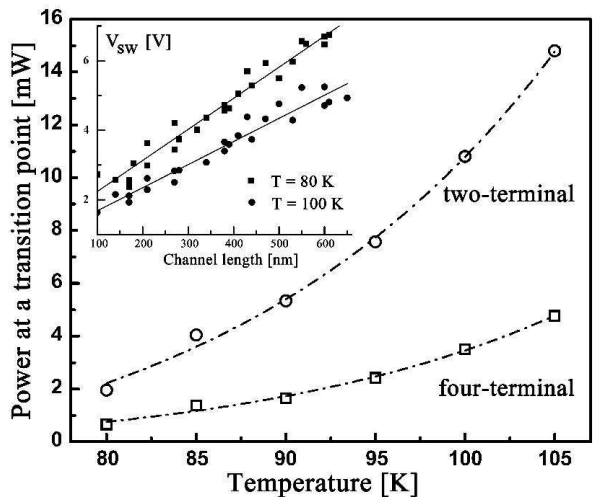}
\end{center}
\vspace{-3mm}
\caption{\small Temperature dependence of the power at a transition point, $V^{On}_{sw}$, in two-terminal ($P^{2T}_{sw}=V\times I$) and four-terminal ($P^{4T}_{sw}= \Delta V\times I$) experiments. Inset shows the dependence of the two-terminal switching voltage, $V_{sw}$, on the channel length at two different temperatures.  Solid lines represent a linear fit; the slope of each line reflects the electric field strength required to drive the transition.}
\label{fig6}
\vspace{-3mm}
\end{figure}

Since heating effects in the On state are responsible for hysteresis in this system, it is worthwhile to review the arguments that the high resistance to low resistance transition itself is NOT driven by heating the sample above Verwey temperature, but indeed is driven by electric field \cite{Our_magnetite_2007}.
First, a strong argument against heating effect is the temperature dependence of the power, $P_{sw}$, dissipated at the transition point.  It was shown in Ref.~\cite{Our_magnetite_2007} for two-terminal devices that $P^{2T}_{sw}=V\times I$ at  $V^{On}_{sw}$ decreases as temperature decreases.  This behavior is incompatible with the assumption that raising of the channel temperature above Verwey temperature (here $T_{V}\sim$108~K) is responsible of the resistance switching.  Assuming a heating effect, the lower the set temperature the more power should be provided to warm up a channel above $T_{V}$, the opposite of the trend observed in the experiment.  In this work we confirm the $P_{sw}$ vs $T$ dependence in the two-terminal geometry (Fig.~\ref{fig6}).  Additionally, we performed four-terminal measurements (details to be given separately \cite{manuscript_in_preparation}) and calculated the corresponding local switching power dissipated in the channel, $P^{4T}_{sw}= \Delta V\times I$  at a transition point, where $\Delta V$ is a voltage difference between two voltage probes within the channel.  $P^{4T}_{sw}$ is much lower than $P^{2T}_{sw}$ because of the significant contact voltage drop at the electrode/Fe$_{3}$O$_{4}$ interface.  The power dissipated directly in the channel, $P^{4T}_{sw}$, shows the same temperature dependence as $P^{2T}_{sw}$ (Fig.~\ref{fig6}), arguing further against simple heating driving the transition.

A second argument for that the transition is electrically driven is the dependence of $V_{sw}$ on the length of the channel.  At each temperature $V_{sw}$ scales {\it linearly} with the channel length, $L$ (Fig.~\ref{fig6} inset).  This implies that at each temperature there is a certain electric field necessary to drive a transition.  Extracted from the slope of $V_{sw}$ vs. $L$, the critical electric field value is about 7$\times 10^{4}$ V/cm at 105 K and it increases slightly up to $\sim$10$^{5}$ V/cm down to 80 K (Fig.~\ref{fig6} inset). These values are much lower than the catastrophic dielectric breakdown field ($\sim$10$^{7}$ V/cm) \cite{diel_break} for most insulators.

Third, pulse experiments that widely span heating conditions show no shift in $V_{sw}$ position upon varying either pulse width or pulse period.  Longer pulses deliver more energy to the channel ($=I \times V \times t$, where $t$ is the duration (width) of the pulse).  Assuming a thermal switching mechanism, the switching would be expected at lower voltages for longer pulses, which is not the case in the experiments.  Different pulse widths and periods have tremendous effects (both in $V^{Off}_{sw}$ positions and current values) on the behavior of the system in the On state (Fig.~\ref{fig2}) when heating is clearly an issue.   In the Off state only minor changes in $I$-$V$ curves were detectable for sweeps with all available widths and periods.  Finally, sourcing a constant voltage for minutes right below $V_{sw}$ value at a certain temperature does not induce or alter the transition.

In conclusion, resistive switching in magnetite thin films from a high resistance state to a low resistance state is driven by electric field, while the hysteresis observed in such switching is a result of thermal effects.  We find that a simple relaxation time model of the thermal processes can describe the shifting of $V^{off}_{sw}$ positions for sweeps with different pulse parameters.  Model parameters indicate that the substrate and electrodes are important in determining the thermal dynamics of the system.  The electric field-driven switching observed in magnetite is different than several mechanisms suggested for RS in perovskites, such as dynamics of oxygen vacancies and interfacial effects.  First, the lack of intrinsic hysteresis differentiates magnetite RS from that in the perovskites.   Second, magnetite RS can be induced only {\it below} the Verwey temperature, $T_V$, whereas RS in perovskites is observed in a wide range of temperatures\cite{2006_Nature_SrTiO3_Waser}. Below $T_V$ magnetite is in a correlated ordered state \cite{2006_PRL_resonant_X_ray_CO, 2005_review_Wright_CO}.  Thus, RS in Fe$_3$O$_4$ is a strong candidate for the theoretically predicted breakdown of {\it charge-ordered} states by electric field \cite{2008_PRB_theory}.  The mechanism of this nonequilibrium transition demands further investigation.

This work was supported by the US Department of Energy grant
DE-FG02-06ER46337.  DN also acknowledges the David and Lucille Packard
Foundation and the Research Corporation. RGSS and
IVS acknowledge the Science Foundation of Ireland grant 06/IN.1/I91.





\begin{thebibliography}{10}

\bibitem{2008_Sawa_review}
A.~Sawa,
\newblock Mater. Today {\bf 11}, 28 (2008).

\bibitem{2007_Aono_review}
R.~Waser and M.~Aono,
\newblock Nat. Mater. {\bf 6}, 833 (2007).

\bibitem{2007_APL_organic_RS}
R.~M\"{u}ller, R.~Naulaerts, J.~Billen, J.~Genoe, and P.~Heremans,
\newblock Appl. Phys. Lett. {\bf 90}, 063503 (2007).

\bibitem{1997_Tokura_first}
A.~Asamitsu, Y.~Tomioka, H.~Kuwahara, and Y.~Tokura,
\newblock Nature {\bf 388}, 50 (1997).

\bibitem{2006_Nature_SrTiO3_Waser}
K.~Szot, W.~Speier, G.~Bihlmayer, and R.~Waser,
\newblock Nat. Mater. {\bf 5}, 312 (2006).

\bibitem{2000_PRB_Tokura_cuprate}
Y.~Taguchi, T.~Matsumoto, and Y.~Tokura,
\newblock Phys. Rev. B {\bf 62}, 7015 (2000).

\bibitem{2006_APL_NiO_RS}
D.~C. Kim, S.~Seo, S.~E.~Ahn, D.-S.~Suh, M.~J.~Lee, B.-H.~Park, I.~K.~Yoo, I.~G.~Baek, H.-J.~Kim, E.~K.~Yim, J.~E.~Lee, S.~O.~Park, H.~S.~Kim, U-In~Chung, J.~T.~Moon, and B. I. Ryu,
\newblock Appl. Phys. Lett. {\bf 88}, 202102 (2006).

\bibitem{2005_JAP_TiO2_Waser}
B.~J. Choi, D.~S.~Jeong, S.~K.~Kim, C.~Rohde, S.~Choi, J.~H.~Oh, H.~J.~Kim, C.~S.~Hwang, K.~Szot, R.~Waser, B.~Reichenberg, and S.~Tiedke,
\newblock J. Appl. Phys. {\bf 98}, 033715 (2005).

\bibitem{2008_Sun_APL}
C.~Li, X.~Zhang, Z.~Cheng, and Y.~Sun, 
\newblock Appl. Phys. Lett. {\bf 93}, 152103 (2008).

\bibitem{2008_Li_EPL}
L.~J. Zeng, H.~X. Yang, Y.~Zhang, H.~F. Tian, C.~Ma, Y.~B. Qin, Y.~G. Zhao, and J.~Q. Li, 
\newblock EPL {\bf 84}, 57011 (2008).

\bibitem{Our_magnetite_2007}
S.~Lee, A.~Fursina, J.~T.~Mayo, C.~T.~Yavuz, V.~ L.~Colvin, R.~G.~S.~Sofin, I.~V.~Shvets, and D.~Natelson,
\newblock Nature Mater. {\bf 7}, 130 (2007).

\bibitem{2008_APL_HAR_Cr}
A.~Fursina, S.~Lee, R.~Sofin, I.~Shvets, and D.~Natelson,
\newblock Appl. Phys. Lett. {\bf 92} (2008).

\bibitem{1939_Verwey_first_Nature}
E.~Verwey,
\newblock Nature {\bf 144}, 327 (1939).

\bibitem{1969_Tinduced_MIT_Fe3O4_1}
P.~J. Freud and A.~Z. Hed,
\newblock Phys. Rev. Lett. {\bf 23}, 1440 (1969).

\bibitem{1969_Tinduced_MIT_Fe3O4_2}
T.~Burch, P.~P.~Craig, C.~Hedrick, T.~A.~Kitchens, J.~I.~Budnick, J.~A.~Cannon, M.~Lipsicas, and D.~Mattis,
\newblock Phys. Rev. Lett. {\bf 23}, 1444 (1969).

\bibitem{2008_Shvets_growth}
S.~K.~Arora, H.-C.~Wu, H.~Yao, W.Y.~Ching, R.~J.~Choudhary, I.~V.~Shvets, and O.~N.~Mryasov,
\newblock IEEE Trans. Magn. {\bf 44}, 2628 (2008).

\bibitem{2004_JAP_Schvets}
Y.~Zhou, X.~Jin, and I.~V. Shvets,
\newblock J. Appl. Phys {\bf 95}, 7357 (2004).

\bibitem{Shvets_backscat}
A.~Koblischka-Veneva, M.~R.~Koblischka, Y.~Zhou, S.~Murphy, F.~Mu\"{u}cklich, U.~Hartmann, and I.~V.~Shvets,
\newblock J. Magn. Magn. Mater. {\bf 316}, 663 (2007).

\bibitem{Shvets_high_res_Xray}
S.~K. Arora, R.~G.~S. Sofin, I.~V. Shvets, and M.~Luysberg,
\newblock J. Appl. Phys. {\bf 100}, 073908 (2006).

\bibitem{heat_trasf_book}
S.~Volz,
\newblock {\em Microscale and Nanoscale Heat Transfer},
\newblock (Springer-Verlag Berlin Heidelberg, 2007).

\bibitem{1985_heat_capacit}
J.~P. Shepherd, J.~W. Koenitzer, R.~Arag\'on, C.~J. Sandberg, and J.~M. Honig,
\newblock Phys. Rev. B {\bf 31}, 1107 (1985).

\bibitem{manuscript_in_preparation}
A.~A.~Fursina, R.~G.~S.~Sofin, I.~V.~Shvets, and D.~Natelson,
\newblock manuscript in preparation  (2009).

\bibitem{diel_break}
J.~McPherson, J.~Kim, A.~Shanware, H.~Mogul, and J.~Rodriguez,
\newblock IEEE Trans. Electron Devices {\bf 50}, 1771 (2003).

\bibitem{2006_PRL_resonant_X_ray_CO}
D.~J. Huang, H.-J.~Lin, J.~Okamoto, K.~S.~Chao, H.-T.~Jeng, G.~Y.~Guo, C.-H.~Hsu, C.-M.~Huang, D.~C.~Ling, W.~B.~Wu, C.~S.~Yang, and C.~T.~Chen,
\newblock Phys. Rev. Lett. {\bf 96}, 096401 (2006).

\bibitem{2005_review_Wright_CO}
R.~J. Goff, J.~P. Wright, J.~P. Attfield, and P.~G. Radaelli,
\newblock J. Phys.: Condens. Matter {\bf 17}, 7633 (2005).

\bibitem{2008_PRB_theory}
N.~Sugimoto, S.~Onoda, and N.~Nagaosa,
\newblock Phys. Rev. B {\bf 78}, 155104 (2008).

\end{thebibliography}

\end{document}